\documentclass[a4paper]{article}

\usepackage{INTERSPEECH2018}

\title{Deep Extractor Network for Target Speaker Recovery From Single Channel Speech Mixtures}
\name{Jun Wang$^1$, Jie Chen$^1$, Dan Su$^1$, Lianwu Chen$^1$, Meng Yu$^1$, Yanmin Qian$^3$ \thanks{$^3$Yanmin Qian did this work when he was a consultant in Tencent}, Dong Yu$^2$}
\address{
  $^1$Tencent AI Lab, Shenzhen, China\\
  $^2$Tencent AI Lab, Bellevue WA, USA\\
  $^3$Department of Computer Science and Engineering, Shanghai Jiao Tong University, Shanghai, China}
\email{\{joinerwang, leojiechen, dansu, lianwuchen, raymondmyu, yanminqian, dyu\}@tencent.com}

\begin{document}

\maketitle
%
%
%
\begin{abstract}
Speaker-aware source separation methods are promising workarounds for major difficulties such as arbitrary source permutation and unknown number of sources. However, it remains challenging to achieve satisfying performance provided a very short available target speaker utterance (anchor). Here we present a novel "deep extractor network" which creates an extractor point for the target speaker in a canonical high dimensional embedding space, and pulls together the time-frequency bins corresponding to the target speaker. The proposed model is different from prior works in that the canonical embedding space encodes knowledges of both the anchor and the mixture during an end-to-end training phase: First, embeddings for the anchor and mixture speech are separately constructed in a primary embedding space, and then combined as an input to feed-forward layers to transform to a canonical embedding space which we discover more stable than the primary one. Experimental results show that given a very short utterance, the proposed model can efficiently recover high quality target speech from a mixture, which outperforms various baseline models, with \(5.2\%\) and \(6.6\%\) relative improvements in SDR and PESQ respectively compared with a baseline oracle deep attracor model. Meanwhile, we show it can be generalized well to more than one interfering speaker.
\end{abstract}

\section{Introduction}
\label{sec:Instroduction}
Despite great successes brought by deep learning techniques, automatic speech recognition (ASR) systems still perform poorly when the speech is corrupted by interfering speech \cite{DONG-2017, ANA-2015, Qian18}. Given a monaural speech mixture, the problem of recovering sources is intrinsically ill-conditioned as there are an infinite number of possible combinations that are legitimate solutions \cite{PSH-2015}. This results in two major difficulties including arbitrary source permutation and unknown number of sources. Recently, many efforts \cite{YDO-2017, CWE-2015, ANA-2015} have been made to address these difficulties. Permutation invariant training (PIT) \cite{YDO-2017, DONG-2017-1, ZCH2017-2, MKO-2017, YQI-2017} mitigates the permutation problem at the training stage by modifying the training objective function such that labels are permuted to find the closest match to the output of the deep neural network (DNN). Related novel models such as Deep clustering (DC) \cite{JRH-2016,ISIK-2016} and deep attractor networks (DANet) \cite{ZCH-2017} have been proposed to generate discriminative embeddings for each time-frequency (T-F) bin with points from the same source forced to be closer to each other or to an attractor.

With the explosion of applications for speech-driven smart devices, there have been increasing studies on target speaker enhancement \cite{DWA-2017,KZM-2017,MDE-2016}, extraction \cite{KZM2017-2, XLZ-2016}, adaptation \cite{KING-2017,KV-2016} and joint learning with acoustic models\cite{ANA-2015,Heym-2017}. In contrast to the speaker-independent speech separation models, deep learning models for target speaker enhancement or extraction are optimized to predict a single target speaker from the background or speech mixture, hence avoid the difficulties of permutation and unknown source numbers. Conventional approaches extract auxiliary features of the target speaker and concatenate them to the input of a DNN-based acoustic models \cite{KV-2016, Maas-2016, KING-2017}. These approaches can be summarized as speaker-specific network, speaker-specific bias, speaker-specific layer methods. However as pointed out by \cite{KZM-2017}, these approaches generally suffer from deficiencies such as being not extendable to open speaker conditions or not effective enough in capturing speaker characteristics by simply bias adaptation of the input layer. Further, speaker-adaptive layer methods are proposed within deep learning based beam-former framework \cite{DWA-2017,KZM-2017,MDE-2016}, where adaptation utterance of a target speaker is employed to inform a neural network to estimates a T-F mask. The length of adaptation utterance is usually required long enough to capture the target speaker characteristics, e.g., about 10 seconds on average in \cite{KZM2017-2}. Nevertheless, typical application scenarios usually only permit a much shorter adaptation utterance, e.g., a wakeup word for a home assistant device \cite{Maas-2016, KING-2017}. Therefore, a model is hardly feasible nor practical for many realistic scenarios unless it is able to learn speaker characteristics with a very short available adaptation utterance. 

So far no study reveals a human auditory system performs in distinct speaker-dependent or speaker-independent manners separately as discussed above. In contrast, psychoacoustic study in \cite{BRON-2015} suggests an important element of human auditory models for early processing and selection of multi-talker speech is an attentional control initiating a feedback loop, inducing enhancement of to-be-selected input. Inspired by the same principle, we propose a model named as "deep extractor network" (DENet) which exploits a short target speaker utterance that serves as an anchor for attentional control, and constructs a canonical high dimensional embedding space where an extractor is formed to pull together the T-F bins of the to-be-selected input. Although the proposed model discussed in this paper still requires a speaker-aware input as an anchor, our single unified framework has the flexibility to extend the construction of anchor embeddings to speaker-independent ways such as a feedback loop.

The rest of the paper is structured as follows. Section 2 gives a brief review of closely related work, and Section 3 describes our proposed method in detail. In Section 4, we describe the experiments and have a discussion based on the results.

\section{Related Work}
\label{sec:DAN}

Our work is mostly related to the DANet \cite{ZCH-2017} which uses a deep learning framework to solve a speaker-independent single channel speech separation problem. Among various cutting-edge speech separation methods, choosing the DANet as our baseline infrastructure is motivated by the fact that it follows the well-studied effects in human speech perception called "perceptual magnet effects" \cite{PAT-1991} which suggest that the brain circuits create perceptual attractors (magnets) that warp the stimulus space such that to draw the sound that is closest to it, and similarly forms an attractor point for each source in the embedding space which draws all the T-F bins towards itself, and using the similarity between the embedded points and each attractor, a mask is estimated for each source in the mixture. The DANet is then trained to minimize the reconstruction error of each source by optimizing the embeddings.

During test time of the DANet, the attractor points for inference are formed using post K-means algorithms, or using an alternative method by finding centriod locations of all attractors of training data. Although the DANet does not have a hard constraint on the number of speakers in the mixtures, the clustering step requires knowing or estimating the number of speakers.

\section{Models}
\label{sec:Models}

Different from speaker separation task of the DANet, our task in this paper is to extract the target speaker from the speech mixture, given a short anchor. In Section \ref{sec:DEN}, we present the model structure of our proposed DENet. To demonstrate its effectiveness, variations of DANet models are then presented which exploit anchor information in alternative ablation ways: In Section \ref{sec:Anchor}, an anchor based DANet uses only anchor to construct attractors in the conventional embedding space; In Section \ref{sec:Nearest}, a nearest attractor based DANet uses anchor only for inference.

Models in Section \ref{sec:DEN} and \ref{sec:Anchor} are trained to minimize the reconstruction error between a masked signal and a clean target reference, such that it minimizes the following objective function:
\begin{equation}
  \boldsymbol{L}=\sum_{f,t}\left \| S_{f,t}^{s}-X_{f,t}\times M_{f,t}^{s} \right \|_{2}^{2}
  \label{eq4}
\end{equation}
where \( S \) is the spectrogram ( frequency \( F  \times \) time \( T \) ) of the target speaker ( we use superscript "\(s\)" to denote the target speaker's speech). \( X \) is the mixture spectrogram ( frequency \( F  \times \) time \( T \) ), and \( M \) is the mask formed to extract the target source. 
A standard L2 reconstruction error is used in the objective function. Gradient is generated to reduce the global reconstruction error for better target speech extraction, as the reconstruction error reflects the difference between the masked signal and the clean target reference.

\begin{figure}[t]
  \centering
  \includegraphics[width=\linewidth]{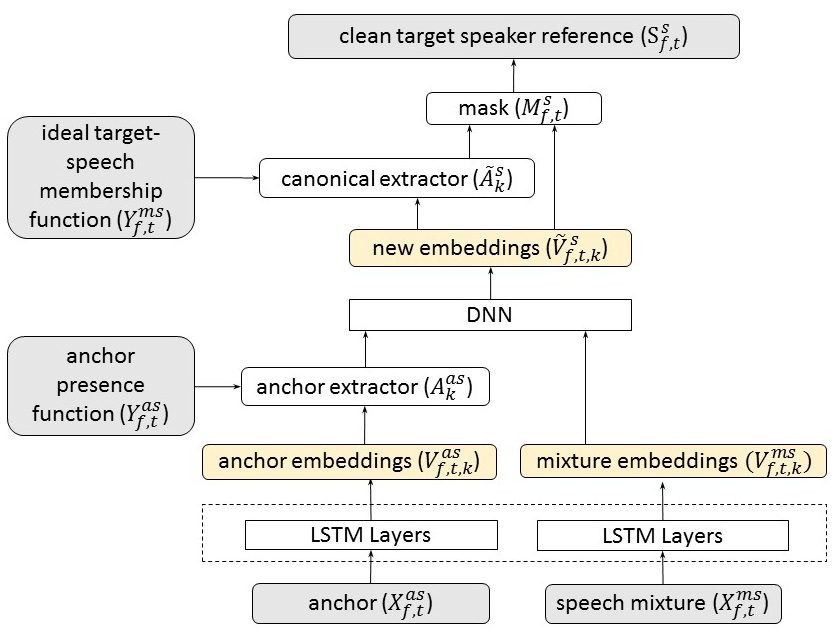}
  \caption{Schematic diagram of "deep extractor network".}
  \label{fig:deep extractor network}
\end{figure}

\subsection{Deep Extractor Network}
\label{sec:DEN}
 
\subsubsection{Training}
\label{sec:Training}
The DENet maps the input signal \( X \) to a canonical \(K\) dimensional embedding space. Its structure is shown in Figure \ref{fig:deep extractor network}: First, through LSTM layers, inputs from an anchor (which we denote with superscript "\(as\)" in the following equations) and a speech mixture (which we denote with superscript "\(ms\)") are firstly mapped to a primary \(K\) dimensional embedding space for each T-F bin, represented by \( \mathbf{V}\in \mathbb{R}^{FT \times K}\).
An extractor of the anchor is calculated using Equation \ref{eq5}:
\begin{equation}
 A_{k}^{as}=\frac{\sum_{f,t}V_{f,t,k}^{as}\times Y_{f,t}^{as}}{\sum_{f,t}Y_{f,t}^{as}}
  \label{eq5}
\end{equation}
where  \( \mathbf{Y}^{as}\in \mathbb{R}^{FT} \) is an anchor presence function for each T-F bin, i.e., \(Y_{f,t}^{as} = 1\) if the amplitude is greater than 40dB below the maximum. 

The extractor is then concatenated with the speech mixture embedding to form an extended embedding of \(2K\) dimension for each T-F bin. The extended embeddings are then fed into a feed forward network to generate new embeddings:
\begin{equation}
  \widetilde{V}_{f,t,k}^{s} =f([{A}_{k}^{as},  V_{f,t,k}^{ms}])
  \label{eq7}
\end{equation}
where \(f(\cdot )\) represents a complex nonlinear mapping function learnt by the deep neural network (DNN), which is introduced to map the original embedding space to a new canonical embedding space represented by \( \widetilde{\mathbf{V}}\in \mathbb{R}^{FT \times K}\).

A new canonical extractor is then estimated based on the new embeddings \(\widetilde{V}_{f,t,k}^{s}\) and the ideal target-speech membership function \(Y_{f,t}^{ms}\):
\begin{equation}
  \widetilde{A}_{k}^{s}=\frac{\sum_{f,t}\widetilde{V}_{f,t,k}^{s}\times Y_{f,t}^{ms}}{\sum_{f,t}Y_{f,t}^{ms}}
  \label{eq8}
\end{equation}

Finally, the mask is estimated in the new embedding space:
\begin{equation}
  M_{f,t}^{s}=Sigmoid(\sum _{k}{\widetilde{A}_{k}^{s}\times \widetilde{V}_{f,t,k}^{s}})
  \label{eq9}
\end{equation}

Equation \ref{eq9} uses a similarity metric defined as a distance between positions of an embedding vector and the extractor in the canonical embedding space. This similarity metric uses the inner product followed by a sigmoid function which monotonically scales the masks between \([0; 1]\). Intuitively, the closer a T-F bin's embedding vector is to the extractor, the higher weights it is to be masked with and assigned as the target speech. 

\subsubsection{Inference}
\label{sec:Inference}
In testing phase, since the true assignment about the ideal target-speech membership function is unknown, a preset extractor is calculated by averaging extractors of all training data, and then used for computing the reconstruction mask as in Equation \ref{eq9}. This method is valid given that the distribution of the extractors are relatively close and stable, which is validated in our experiments.

\subsection{Anchor Based Deep Attractor Network}
\label{sec:Anchor}

This network is based on the conventional DANet \cite{ZCH-2017}. To inform the DANet with the anchor knowledge, intuitively a target speaker attractor point can be estimated using the anchor in Equation \ref{eq5}.

Next, a reconstruction mask for the mixture part of the target speaker is estimated in Equation \ref{eq6} by finding the similarity of each T-F bin in the embedding space to the anchor attractor point.
\begin{equation}
  M_{f,t}^{s}=Sigmoid(\sum _{k}{A_{k}^{as}\times V_{f,t,k}^{ms}})
  \label{eq6}
\end{equation}

In both training and testing phase, the target speaker's attractor is generated using the anchor. A rational assumption is that if an embedding of a T-F bin of the mixture is close to the attractor generated from the anchor, then it is more probable to belong to the target speaker, and the resulting mask for the target speaker will produce larger values for that T-F Bin. 

\subsection{Nearest Attractor Based Network}
\label{sec:Nearest}
An alternative scheme for estimating the target speaker's attractor is to directly use the attractors generated by the DANet from the speech mixture. Hence the training phase follows the routine as the conventional DANet, and a fixed pair of attractors are collected from the training data. 

For reference, since the conventional DANet doesn't have any knowledge about which attractor belongs to the target speaker, we use anchor knowledge for picking the target attractor. Hence for inference, we generate a reference attractor based on the anchor, compare the distances from the fixed attractor pair to the anchor attractor, and then select the one with the smaller distance as the target speaker's attractor.

\subsection{Comparison Of Different Networks}
\label{sec:Relation}
The proposed DENet framework in Section \ref{sec:DEN} has two distinct advantages: 
First, it constructs a single unified network to fully explore both the anchor information and the supervised label of the speech mixtures during the end-to-end training phase. In contrast, the similarity metric in DANet is defined as a distance between absolute positions of an embedding vector and an attractor in the primary embedding space. As observed and pointed out in \cite{ZCH-2017}, more than one attractor pair, dispersing from each other, may have been learnt in the conventional DANet embedding space. This suggests that using absolute positions for the similarity metric could be less stable. Comparatively, the canonical extractor generated in our new framework is expected to capture information from both the anchor and the mixtures from the primary embedding space such that their relative position information is represented in the canonical embedding space. We discover that extractors in the canonical embedding space are more stable than the conventional ones using absolute positions. Secondly, compared to using post K-means algorithm for inference, a fixed preset extractor method has the advantage of real-time implementation using a frame-by-frame pipeline. Moreover, since the extractor in the canonical embedding space is more stable, it is able to generalize better to unseen speakers and mixtures and to produce better quality separation than using a preset extractor calculated in the primary embedding space. 

\section{Experiments}
\label{sec:Exp}
\subsection{Data}
\label{sec:Data}
Our corpus contained \(373\) speakers in all, and for each speaker, there were about \(100\) anchor speech samples and \(500\) speech mixture samples. The speech mixtures included frequently used commands or queries consisting of \(2\) to \(22\) Chinese characters. The average length of the anchor samples was \(0.9\) seconds, and that of the speech mixtures was \(2.3\) seconds. 
All speech data were recorded with \(16\)Bit, \(16\)kHz and single-channel format in moderately reverberant environments. Short-time Fourier transform (STFT) magnitude was computed as the input feature, with \(32\)ms window length, \(16\)ms hop size, and the square root of Hann window.

We randomly selected \(323\) speakers as a training set and used the remaining \(50\) speakers for testing. For both the training set and testing set, the speech mixtures were generated with signal-to-interference ratio (SIR) ranging from \(0\) to \(10\) dB. The interference speakers were randomly selected from a different corpus, which had no overlap with the \(373\) speakers. 
\subsection{Experimental Setup}
\label{sec:Setup}

\subsubsection{Baseline Models}
\label{sec:Baseline}

The first baseline model was an oracle DANet which we denoted as \(DANet\ Oracle\): To get a glimpse of the upper bound of target speech quality separated by using the conventional DANet, the two speakers separated by the DANet were compared with our ground-truth target speaker, and the one with the best SDR was selected as the extracted target. The second baseline model was the anchor based DANet as described in Section \ref{sec:Anchor} which we denoted as \(DANet\ Anchor\), and the third baseline model was the nearest attractor based network as described in Section \ref{sec:Nearest} which we denoted as \(DANet\ Nearest\).

We constructed the above three baseline models with the same configuration of the DANet as \cite{ZCH-2017}, which consisted of \(4\) Bi-directional LSTM layers with \(600\) hidden units each layer. In our experiments, the embedding dimension was set to \(40\), resulting in a fully connected feed-forward layer of \(10280\) hidden units \( ( 40 \times 257 ) \) after the BLSTM layers.

\subsubsection{Deep Extractor Network}
The LSTM layers were configured with the same setting as the DANet. For the DNN structure, we used a \(2\)-layer feed-forward network, where the input dimension was \(80\) (i.e. double of the DANet embedding size), the hidden layer had \(256\) units, and the output dimension was \(40\). 
The parameter size of the proposed DENet was $279,797K$, which was comparable to the parameter size $279,425K$ of the baseline models, as the increased size by DNN was very small and neglectable.

For all the above models, to reduce noisy low-amplitude T-F bins, those with amplitudes \(40\)dB below the maximum were ruled out for calculating extractors. For training \(DANet\ Oracle\) and \(DANet\ Nearest\), we followed the conventional DANet to apply a curriculum training strategy that we first trained the network with \(100\)-frame input length and then continued training with \(400\)-frame input length, whereas for \(DANet\ Anchor\) and \(DENet\), the network were trained with utterance-level input.

\subsection{Results}
 
Table \ref{tab:two_speaker} showed the signal-to-distortion ratio (SDR) \cite{Vincent06} and Perceptual Evaluation of Speech Quality (PESQ) \cite{Rix01} results for different models. The first model denoted as \(Orig\ Mixture \) meant the unprocessed mixture speech. As shown in the result, all models gained large performance boost compared with this reference. The proposed \(DENet\) brought the best performance in both SDR and PESQ, with a \(2.3\%\) relative improvement in SDR compared with \(DANet\ Anchor\) , and a further \(5.2\%\) relative improvement compared with \(DANet\ Oracle\).

\begin{table}[th]
  \caption{SDR and PESQ results for different models.}
  \label{tab:two_speaker}
  \centering
  \begin{tabular}{ c c c }
    \toprule
    Model &  SDR &  PESQ  \\
    \midrule
    $Orig\ Mixture$          & $0.99$   & $1.96$ \\
    $DANet\ Nearest$          & $15.71$  & $2.51$ \\
    $DANet\ Oracle$               & $16.67$  & $2.58$ \\
    $DANet\ Anchor$    & $17.14$  & $2.72$ \\
    $\textbf{DENet}$     & $\textbf{17.53}$  & $\textbf{2.75}$ \\

    \bottomrule
  \end{tabular}

\end{table}

It was also worth noting that \(DENet\) and \(DANet\ Anchor\) both significantly outperformed \(DANet\ Oracle\) and \(DANet\ Nearest\), suggesting that a training phase using a reliable extractor estimation via anchor was important to achieve good performances. Also we stressed that the average length of anchor in our dataset was \(0.9\)s, which was remarkably shorter than the length required (about \(10\)s) in state-of-the-art deep learning based models which employed adaptation utterance of a target speaker. This revealed that speaker characteristics could be successfully learnt by training models with speaker adaptation in an extractor/ attractor framework, despite of a very short adaptation utterance, which could be crucial for a successful application in realistic scenarios.

\subsubsection{Analysis of Canonical Extractor}
\begin{figure}[t]
  \centering
  \includegraphics[width=\linewidth]{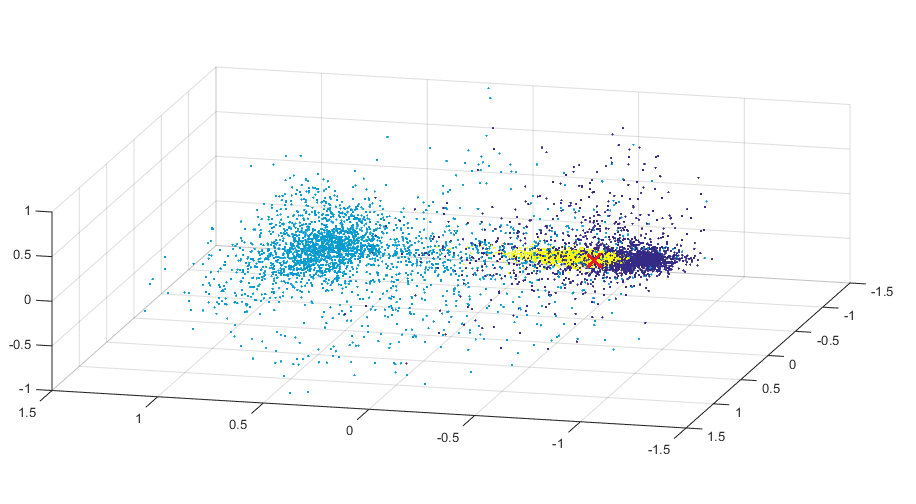}
  \caption{Location of extractors and embeddings of T-F bins in the new embedding space. Each dot visualizes the first three principle components; Yellow dots indicate canonical extractors of all training speakers, and their centroid is marked with a red cross; Light blue dots indicate embeddings for an interfering speaker, and dark blue dots for a target speaker.}
  \label{fig:new_embedding_space}
\end{figure}

Figure \ref{fig:new_embedding_space} showed extractors and embeddings of T-F bins in the new embedding space. For visualization purposes, the first three principle components were plotted. Yellow dots represented canonical extractors of all speakers in the training set, and their centroid was marked with a red cross. For simple display, we presented embeddings of a random speech mixture sample, which were typical and representative among different mixture samples: light blue dots represented embeddings for the interfering speaker, and dark blue dots for the target speaker. It could be observed that embeddings of the interfering speaker were distant from the canonical extractors, while those of the target speaker were much closer to and actually wrapping the canonical extractors, thus leading to a high-quality target speech recovery. We could also observe a good property of the canonical extractors that these yellow dots for different speakers were very stable and closely clustered. This observation echoed our discussion in Section \ref{sec:Relation} that extractors in the canonical embedding space encoding relative information were more stable than the conventional ones \cite{ZCH-2017} using absolute information.

\subsubsection{Generalization to More Interfering Speakers}
Since the proposed DENet model required no prior knowledge about the number of speakers, it could be flexibly extended to cope with more than one interfering speaker. We ran experiments on different models to evaluate their generalization and robustness performances on three-speaker mixtures. Note that all models in this experiment were trained only with two speaker mixtures. 

\begin{table}[th]
  \caption{SDR and PESQ results for target speaker extraction from three-speaker mixtures in medium (left columns) and hostile (right columns) interfering conditions}
  \label{tab:three_speaker}
  \centering
  \begin{tabular}{ l c c | c c }
    \toprule
    Model & SDR & PESQ & SDR & PESQ\\
    \midrule
    $Orig\ Mix.$    & $0.44$   & $1.87$ & $-2.05    $ & $1.46$  \\
    $DANet\ Near.$  & $11.99$   & $2.30$        & $8.84\ (26\%)$  & $2.13$  \\
    $DANet\ Orac.$     & $11.98$   & $2.27$          & $8.85\ (26\%)$  & $2.10$   \\
    $DANet\ Anch.$  & $13.32$   & $2.48$  & $10.39\ (22\%)$  & $2.12$ \\
    $\textbf{DENet}$  & $\textbf{13.44}$   & $\textbf{2.52}$   & $\textbf{10.67\ (20\%)}$  & $\textbf{2.14}$ \\

    \bottomrule
  \end{tabular}

\end{table}

Two sets of evaluation results were shown in Table \ref{tab:three_speaker}: left columns for a medium interfering condition with $Orig\ Mix.$ at \(SDR=0.44\), and right columns for a hostile interfering condition with $Orig\ Mix.$ at \(SDR=-2.05\). The ratio numbers in parenthesis indicated relative degradations between the hostile and medium conditions. It was shown that the proposed \(DENet\) model consistently outperformed the other models in all conditions for the three-speaker mixture case. 
It could be expected that better results could further be achieved by adding training data with multiple interfering speakers and under more interfering conditions, and we left it for our future work.

\section{Conclusions}
A new unified network called DENet is investigated to extract the target speaker from speech mixtures, given a short anchor utterance. The training process fully explores both anchor information and supervised mixture labels by representing their relative information in a canonical high dimensional embedding space which is shown more stable than a primary embedding space by a conventional DANet. Experimental results show that even with a very short anchor (less than \(1\) second), the proposed DENet is capable of recovering high quality target speech. Ablation experiments have been conducted by comparing the proposed DENet with alternative DANet-based models, and best performances have been consistently achieved by the proposed model under all test conditions. Furthermore, we show that this approach can be generalized well to speech mixtures with more than one interfering speaker.
In our future work, we will explore the DENet for more complex conditions of far-field environments and multiple interferences. We will also explore speaker-independent anchor embedding construction in the DENet via an attentional control and feedback loop mechanism. 

\bibliographystyle{IEEEtran}

\bibliography{mybib}

\begin{thebibliography}{10}
\providecommand{\url}[1]{#1}
\csname url@samestyle\endcsname
\providecommand{\newblock}{\relax}
\providecommand{\bibinfo}[2]{#2}
\providecommand{\BIBentrySTDinterwordspacing}{\spaceskip=0pt\relax}
\providecommand{\BIBentryALTinterwordstretchfactor}{4}
\providecommand{\BIBentryALTinterwordspacing}{\spaceskip=\fontdimen2\font plus
\BIBentryALTinterwordstretchfactor\fontdimen3\font minus
  \fontdimen4\font\relax}
\providecommand{\BIBforeignlanguage}[2]{{%
\expandafter\ifx\csname l@#1\endcsname\relax
\typeout{** WARNING: IEEEtran.bst: No hyphenation pattern has been}%
\typeout{** loaded for the language `#1'. Using the pattern for}%
\typeout{** the default language instead.}%
\else
\language=\csname l@#1\endcsname
\fi
#2}}
\providecommand{\BIBdecl}{\relax}
\BIBdecl

\bibitem{DONG-2017}
D.~Yu and J.~D. Li, ``Recent progresses in deep learning based acoustic
  models,'' \emph{IEEE/CAA JOURNAL OF AUTOMATICA SINICA}, vol.~4, no.~3, 2017.

\bibitem{ANA-2015}
A.~Narayanan and D.~Wang, ``Improving robustness of deep neural network
  acoustic models via speech separation and joint adaptive training,''
  \emph{IEEE/ACM Trans. ASLP}, vol.~23, no.~1, pp. 92--101, 2015.

\bibitem{Qian18}
Y.~Qian, C.~Weng, X.~Chang, S.~Wang, and D.~Yu, ``Past review, current
  progress, and challenges ahead on the cocktail party problem,''
  \emph{Frontiers of Information Technology \& Electronic Engineering},
  vol.~19, no.~1, pp. 40--63, 2018.

\bibitem{PSH-2015}
P.~S. Huang, M.~Kim, M.~H. Johnson, and P.~Smaragdis, ``Joint optimization of
  masks and deep recurrent neural networks for monaural source separation,''
  \emph{IEEE/ACM Trans. Audio, Speech, Lang. Process.}, vol.~23, no.~12, pp.
  2136--2147, 2015.

\bibitem{YDO-2017}
Y.~Dong, K.~Morten, T.~Zheng-Hua, and J.~Jesper, ``Permutation invariant
  training of deep models for speaker-independent multi-talker speech
  separation,'' \emph{ICASSP'17}, pp. 31–--35, 2017.

\bibitem{CWE-2015}
C.~Weng, D.~Yu, M.~L. Seltzer, and J.~Droppo, ``Deep neural networks for
  single-channel multi-talker speech recognition,'' \emph{IEEE Transactions on
  Audio, Speech and Language Processing}, vol.~23, no.~10, pp. 1670--–1679,
  2015.

\bibitem{DONG-2017-1}
D.~Yu, X.~Chang, and Y.~M. Qian, ``Recognizing multi-talker speech with
  permutation invariant training,'' \emph{INTERSPEECH'17}, 2017.

\bibitem{ZCH2017-2}
Z.~Chen, J.~Droppo, J.~Li, and W.~Xiong, ``Progressive joint modeling in
  unsupervised single-channel overlapped speech recognition,'' \emph{arxiv,
  vol. abs/1707.07048}, 2017.

\bibitem{MKO-2017}
M.~Kolbaek, D.~Yu, Z.~H. Tan, and J.~Jensen, ``Joint separation and denoising
  of noisy multi-talker speech using recurrent neural networks and permutation
  invariant training,'' \emph{MLSP'17}, 2017.

\bibitem{YQI-2017}
Y.~Qian, X.~Chang, and D.~Yu, ``Single-channel multi-talker speech recognition
  with permutation invariant training,'' \emph{arxiv, vol. abs/1707.06527},
  2017.

\bibitem{JRH-2016}
J.~R. Hershey, Z.~Chen, J.~L. Roux, and S.~Watanabe, ``Deep clustering:
  Discriminative embeddings for segmentation and separation,''
  \emph{ICASSP'16}, pp. 31–--35, 2016.

\bibitem{ISIK-2016}
Y.~Isik, J.~L. Roux, Z.~Chen, S.~Watanabe, and J.~R. Hershey, ``Single-channel
  multi-speaker separation using deep clustering,'' \emph{INTERSPEECH'16}, pp.
  545–--549, 2016.

\bibitem{ZCH-2017}
Z.~Chen, Y.~Luo, and N.~Mesgarani, ``Deep attractor network for
  single-microphone speaker separation,'' \emph{ICASSP'17}, 2017.

\bibitem{DWA-2017}
D.~Wang and J.~Chen, ``Supervised speech separation based deep learning: An
  overview,'' \emph{arxiv}, 2017.

\bibitem{KZM-2017}
K.~Zmolikova, M.~Delcroix, K.~Kinoshita, T.~Higuchi, A.~Ogawa, and T.~Nakatani,
  ``Speaker-aware neural network based beamformer for speaker extraction in
  speech mixtures,'' \emph{Interspeech'17}, 2017.

\bibitem{MDE-2016}
M.~Delcroix, K.~Kinoshita, C.~Yu, A.~Ogawa, T.~Yoshioka, and T.~Nakatani,
  ``Context adaptive deep neural networks for fast acoustic model adaptation in
  noisy conditions,'' \emph{ICASSP'16}, p. 5270–5274, 2016.

\bibitem{KZM2017-2}
K.~Zmolikova, M.~Delcroix, K.~Kinoshita, T.~Higuchi, A.~Ogawa, and T.~Nakatani,
  ``Learning speaker representation for neural network based multichannel
  speaker extractions,'' \emph{ASRU'17}, Dec, 2017.

\bibitem{XLZ-2016}
X.~L. Zhang and D.~Wang, ``A deep ensemble learning method for monaural speech
  separation,'' \emph{IEEE/ACM Trans. Audio, Speech, Lang. Process.}, vol.~24,
  no.~5, pp. 967--977, 2016.

\bibitem{KING-2017}
B.~King, I.~Chen, Y.~Vaizman, Y.~Liu, R.~Maas, S.~H.~K. Parthasarathi, and
  B.~Hoffmeister, ``Robust speech recognition via anchor word
  representations,'' \emph{INTERSPEECH'17}, 2017.

\bibitem{KV-2016}
K.~Vesely, S.~Watanabe, K.~Zmolikova, M.~Karafiat, L.~Burget, and J.~H.
  Cernocky, ``Sequence summarizing neural network for speaker adaptation,''
  \emph{ICASSP'16}, 2016.

\bibitem{Heym-2017}
J.~Heymann, L.~Drude, C.~Boeddeker, P.~Hanebrink, and R.~Haeb-Umbach,
  ``Beamnet: End-to-end training of a beamformer-supported multi-channel asr
  system,'' \emph{ICASSP'17}, pp. 5325--5329.

\bibitem{Maas-2016}
R.~Maas, S.~H.~K. Parthasarathi, B.~King, R.~Huang, and B.~Hoffmeister,
  ``Anchored speech detection,'' \emph{INTERSPEECH'16}, pp. 2963--2967.

\bibitem{BRON-2015}
A.~W. Bronkhorst, ``The cocktail-party problem revisited: early processing and
  selection of multi-talker speech,'' \emph{Attention, Perception, and
  Psychophysics, Springer}, 2015.

\bibitem{PAT-1991}
P.~K. Kuhl, ``Human adults and human infants show a perceptual magnet effect,''
  \emph{Perception and Psychophysics, Springer}, pp. 93--107, 1991.

\bibitem{Vincent06}
E.~Vincent, R.~Gribonval, and C.~Fevotte, ``Performance measurement in blind
  audio source separation,'' \emph{IEEE/ACM Trans. Audio, Speech, Lang.
  Process.}, vol.~14, no.~4, pp. 1462--1469, 2006.

\bibitem{Rix01}
A.~Rix, J.~Beerends, M.~Hollier, and A.~Hekstra, ``Perceptual evaluation of
  speech quality (pesq)-a new method for speech quality assessment of telephone
  networks and codecs,'' \emph{ICASSP'01}, pp. 749--752.

\end{thebibliography}



%



\end{document}